\documentclass[mathleft]{an}
\usepackage{graphicx}
\usepackage{txfonts}
\usepackage{subfigure}
\usepackage{times}
\begin{document}

% The following seven commands are intended for editorial usage and should be ignored by
% the author(s).
\Pagespan{789}{}% Document's page range. 
% If second parameter is left empty, the last page is computed automatically.
%\Yearpublication{2008}%
%\Yearsubmission{2008}%
\Month{08}%   
\Volume{999}%  
\Issue{88}% 
% \DOI{This.is/not.aDOI}% 

\title{Modeling Merging Galaxies using MINGA -- \\
        Improving Restricted {\it N-}body by Dynamical Friction}

\author{H.P. Petsch \inst{1}\fnmsep\thanks{Corresponding author:
  \email{petsch@astro.univie.ac.at}\newline}
%Example 
%for footnote, note the usage of the \texttt{fnmsep}
%command as separator between institute number and footnote mark} 
\and  Ch. Theis \inst{1}
}
\titlerunning{Modeling Merging Galaxies}
\authorrunning{H.P. Petsch \& Ch. Theis}
\institute{Institut f\"ur Astronomie, Universit\"at Wien, T\"urkenschanzstrasse 17, 1180 Wien, Austria}

%\received{2008}
%\accepted{2008}
%\publonline{later}


%To reduce the computational expenses, advanced simulations require a reliable set of initial conditions for the interaction and the galaxies themselves. 
%For a quantitative confrontation with observations, however, the method has to be adapted to the properties of real galaxies. E.g.\,realistic treatments of the dark matter halos have to be included.
\abstract{Modeling interacting galaxies to reproduce observed systems is still a challenge due to the extended parameter space (among other problems). 
Orbit and basic galaxy parameters can be tackled by fast simulation techniques like the restricted \textit{N}-body method, applied in the fundamental work by Toomre \& Toomre (1972). This approach allows today for the study of millions of models in a short time. One difficulty for the classical restricted \textit{N}-body method is the missing orbital decay, not allowing for galaxy mergers. Here we present an extension of the restricted \textit{N}-body method including dynamical friction. This treatment has been developed by a quantitative comparison with a set of self-consistent merger simulations. By varying the dynamical friction (formalism, strength and direction), we selected the best-fitting parameters for a set of more than 250\,000 simulations. We show that our treatment reliably reproduces the orbital decay and tidal features of merging disk galaxies for mass ratios up to $q=1/3$ between host and satellite. We implemented this technique into our genetic algorithm based modeling code \textsc{Minga} and present first results.}

\maketitle

\section{Introduction}
Finding the initial parameters of an interacting galactic system is still like looking for a needle in a haystack. One challenge is the large number of parameters describing the orbital and the galactic properties. Some of these parameters might be derived from detailed observations, e.g.\,from HI data cubes. However, to perform an effective search in a high dimensional parameter space, it is necessary to use fast simulations and sophisticated finding strategies in parameter space.
%In order to determine all important large scale parameters of the interacting galaxy system, 
We use the code \textsc{Minga} (Theis 1999), where an improved restricted \textit{N-}body code is coupled to a genetic algorithm (GA). Such a strategy has also been proposed by Wahde (1998). For the galaxy NGC\,4449 Theis \& Kohle (2001) showed, that the H\,I structure of a weakly interacting system can be reproduced.

\subsection{Genetic Algorithm}
\textsc{Minga} uses a genetic algorithm based on \emph{pikaia} (Charbonneau 1995). This kind of algorithms try to imitate nature regarding the evolution of species. Heredity and mutation of characteristics are used to adopt the simulations to the observations. Each model parameter is coded (normalised) to a \emph{gene} (here we use 4 digits for a gene). All genes together are then forming a single string, the \emph{chromosome}, which is fully describing a complete interaction model. The realisation of heredity and the determination of the fitness\footnote{Fitness is a quantitative measure of the quality of a model.} can be done using quite different techniques. Here we use a cross-over operator that cuts two chromosomes at a random position and swaps the remaining ends. Better fitting models are more likely \emph{parents} of the next generation of models. This process is used to evolve the models from generation to generation. The fitness of the models is usually raising, especially if elitism\footnote{The best model of a generation is forwarded, if no superior model was found in the next generation.} is used. However, this evolution process could suffer from \emph{inbreeding}\footnote{The optimisation process got stuck caused by a too homogeneous set of individual models.} and therefore mutation\footnote{With a low probability each chromosome entry might be changed.} is applied. \textsc{Minga} uses either constant or changing mutation rates (depending on symbols for inbreeding). A more detailed description of the GA is provided in Theis (1999) and Theis \& Kohle (2001).
% Too high mutation rates could lead in rather random search in parameter space than in model evolution.

\subsection{Restricted \textit{N-}body}
%The gravitative \textit{N-}body-problem is described by the Newtonian equation of motion Eq.\,(\ref{gl_nk_1}).
%{\setlength{\mathindent}{0pt}\begin{equation} \ddot{\vec{r}}_i= \frac{\vec{F}_i}{m_i}=-G\sum_{\stackrel{j=1}{j \neq i}}^N \frac{m_j}{\left|\vec{r}_i-\vec{r}_j\right|^3} %\cdot \left(\vec{r}_i-\vec{r}_j \right) \label{gl_nk_1}
%\end{equation}}
%{$N$ denotes the number of particles, $\vec{r}_i$ the position of the $i$-th particle and $m_i$ its mass. $G$ represents the constant of gravitation.}
The first who applied the restricted \textit{N-}body method to interacting galaxies were Pfleiderer \& Siedentopf (1961) and Toomre \& Toomre (1972) -- hereafter TT72. This approach treats the galactic centres self-consistently, while the disk consists of (mass-free) test particles. The main advantage of the restricted \textit{N}-body method is the reduction of the $O (N^2)$ problem of the original Newtonian equation of motion to about $O(N N_G)$, if $N_G$ denotes the number of galaxies. For point mass galaxies the set of equations is reduced to

{\setlength{\mathindent}{0pt}\begin{equation}
\ddot{\vec{r}}_i=\frac{\vec{F}_i}{m_i}=- G \sum_{k=1}^{N_G} {\frac{M_k}{\left|\vec{r}_i-\vec{R}_k(t)\right|^3}}\cdot \left(\vec{r}_i -\vec{R}_k(t) \right).\label{petsch_eq_1}
\end{equation}}

\noindent $\vec{r}_i$ is the position of the $i$-th particle and $m_i$ its mass. $\vec{R}_k(t)$ describes the position of galaxy $k$ at time $t$ and $M_k$ is its dynamical mass including dark matter. $G$ represents the constant of gravitation.

Different to TT72, \textsc{Minga} allows for a self-consistent description of (rigid) extended halos (Gerds 2001; Theis 2004). Though this treatment substantially influences the galactic orbits (and also increases the CPU time), the restricted \textit{N-}body method, i.e.\,Eq.\,(\ref{petsch_eq_1}), can still be applied.

Another important process is dynamical friction. It describes the deceleration (due to scattering) of a perturber moving in a background of particles. Self-consistent modeling already accounts for dynamical friction, but it is missing in classical restricted \textit{N-}body codes and, therefore, these codes were not able to remodel tightly interacting or merging systems.

\subsection{Dynamical Friction}
%Despite the absence of dissipative forces within pure \textit{N-}body systems, such as stars or Dark Matter particles, interaction leads to redistribution of energy from orbital to internal thus to decaying orbits. 
A simple formula for dynamical friction was derived by Chandrasekhar (1942) by using following assumptions: A point mass perturber is moving in an homogeneous, infinite background of particles and the mass of a background particle is negligible compared to the perturbers mass.

\begin{equation} \frac{d \vec{v}_M}{dt}= -F(v_M,\sigma) \frac{\rho M}{v_M^3} \ln \Lambda \ \vec{v}_M
\label{petsch_eq_2}\end{equation}

The acceleration $d\vec{v}_M / dt$ of a massive particle $M$ depends on the background density $\rho$, the mass of the perturber $M$ and its velocity $v_M$. The acceleration is pointing opposite to the direction of the velocity, hence causing an effective deceleration of the particle. For more details on the function $F(v_M,\sigma)$, refer to Binney \& Tremaine (1987). The Coulomb logarithm $\ln \Lambda$ is the relation between the maximum impact parameter $b_{\mathrm{max}}$ and the impact parameter $b_0$ that leads to a $90^\circ$ degree deflection:
%The most critical parameter within Eq.\,(\ref{petsch_eq_2}) is the Coulomb logarithm -- see Eq.\ref{petsch_eq_3}.
%$ \ln \Lambda$ -- see equation (\ref{petsch_eq_3}).

\begin{equation} \Lambda =  \frac{b_{\mathrm{max}} V_0^2}{G(M+m)} = \frac{b_{\mathrm{max}}}{b_0}.
\label{petsch_eq_3} \end{equation}
 \noindent $V_0$ denotes the velocity of the reduced particle (perturber and one background particle with mass $m$).

Recently, efforts have been made to improve shortcomings
%\footnote{Due to over-tightened assumptions like infinite, homogeneous background or point mass perturbers.} 
of the approach, Eq.\,(\ref{petsch_eq_2}). E.g.\,Hashimoto et al.\,(2003) and Spinnato et al.\,(2003) accounted for finite halo systems. Just \& Pe\~{n}arrubia (2005) focused on the influence of a density gradient. Furthermore Jiang et al.\,(2008) claim, that a mass dependency should be applied to the Coulomb logarithm for merging time scales of dark matter galaxies in a cluster.

%\pagebreak%%%%%%%%%%%%%%%%%

%\section{Aims}
%Our main aim here is to improve \textsc{Minga} by adding dynamical friction to the equation of motion of the galaxy halos. Thus allowing the program to derive the parameters of strongly interacting systems and galaxy merger. This improvement will still keep the fast restricted \textit{N-}body method in order to calculate one single model within a few seconds compared to several minutes of a full self-consistent way. We need to find the best description(s) for dynamical friction and its(their) parameterisation(s).

\section{Method}\label{petsch_sec_method}
We are using a set of self-consistent reference models to determine the appropriate formalism of the dynamical friction. We have chosen isothermal spheres to serve as host halo galaxies. Different satellites -- point masses, isothermal spheres and disk galaxies -- are merged with them. The reference models have been evolved using the \emph{gyrfalcON} tree-code (Dehnen 2000). For an independent simulation we also used a direct code on a \emph{Grape6A - board} (Sugimoto et al.\,1990; Makino et al.\,2003). 65\,000 particles per halo have been used for the self-consistent models. The halo was truncated at $150$\,kpc resulting in a mass of $5.4 \cdot 10^{11} M_\odot$ and a velocity dispersion of $\sigma=62$\,km/s. The deviation between radial decay in our models and the reference models is used as a diagnostics. Detailed results of our studies will be published later. We varied the Coulomb logarithm, the strength and the direction of the dynamical friction force. The force itself is applied in a symmetric way to the equations of motion of the galaxy centres. The varied parameters ($C_f$, $\beta$ and $\ln \Lambda$) are shown in Eq.\,(\ref{petsch_eq_4}).

{\setlength{\mathindent}{0pt}\begin{equation} \frac{d \vec{v}_M}{dt} = - F(v_M,\sigma) C_f \frac{\rho M}{v_M^2} \left(\hat{v}_M \cos{(\beta)}  + \hat{e}_\bot \sin{(\beta)} \right) \ln \Lambda 
\label{petsch_eq_4}
\end{equation}}

\noindent $C_f$ is a simple scaling factor that allows for fitting. As galaxies have density gradients, the force might point not exactly opposite the velocity, therefore we introduce an orthogonal component which is adjustable via $\beta$. Finally, $\ln \Lambda$ is derived by different approaches, these also denote our models:

\begin{itemize}
\item \textbf{Model A} uses a constant Coulomb logarithm.
\item \textbf{Model B} uses a distance-dependent Coulomb logarithm as described by Hashimoto et al.\,(2003):
\begin{equation}
 \ln \Lambda= \ln \left(\frac{r_M}{1.4 b_0}\right)
\label{petsch_eq_5}
\end{equation}
\noindent $r_M$ is the distance satellite -- halo centre.
\item \textbf{Model C} uses an interpolation between two constant Coulomb logarithms (not presented here).
%Model C lassen wir mal aus
%\item \textbf{Model C} uses an interpolation between two constant Coulomb logarithms, one in the centre and one at the limit of the host halo Eq.\,(\ref{gl_C}).
\item \textbf{Model D} uses a mass- and distance-dependent Coulomb logarithm, similar to a description by Jiang et al.\,(2008):
%Modell C lassen wir mal aus
%\begin{equation}
%\ln \Lambda=\ln \Lambda(0) + \frac{\ln \Lambda(r_\mathrm{halo}) - \ln \Lambda(0)}{r_{\mathrm{halo}}} r_M
%\label{gl_C}
%\end{equation}
%
\begin{equation}
 \ln \Lambda=\ln \left[1 + \frac{M_{\mathrm{halo}}(r_M)}{M} \right]
\label{petsch_eq_6}
\end{equation}
\noindent $M$ is the mass of the satellite and $M_{\mathrm{halo}}(r_M)$ is the mass of the host halo enclosed within the actual satellite's radius $r_M$.
\end{itemize}

\section{Results}

%\subsection{Point mass satellite}
%Some tests with point mass satellites have been performed. Circular and elliptic initial conditions have been tested over a mass range of $q=1/100$ to $q=1/3$. Models A to D have been applied and it was found, that for mass ratios up to $q=1/20$ a constant (model A) or distance-dependent (model B) Coulomb logarithm is sufficient to describe the whole merging process with a deviation\footnote{$\delta_d$ is derived by integrating the quadratic difference between the compared radial decay curves over time.} of $\delta_d (t_{\mathrm{merge}}) \lesssim 5 \cdot 10^{-2}$. For elliptic initial conditions it was necessary to implement the orientation correction $\beta_\mathrm{max} = \pi /4$, while the circular case could not be improved by this. Form more massive satellites like $q=1/3$ it was necessary to switch to mass-dependent Coulomb logarithm (model D). One complete revolution of the satellite could be reproduced then with an error of $\delta_d (t_\mathrm{revolution}) \lesssim 2 \cdot 10^{-2}$.

\begin{figure} \centering
\includegraphics[width=6cm]{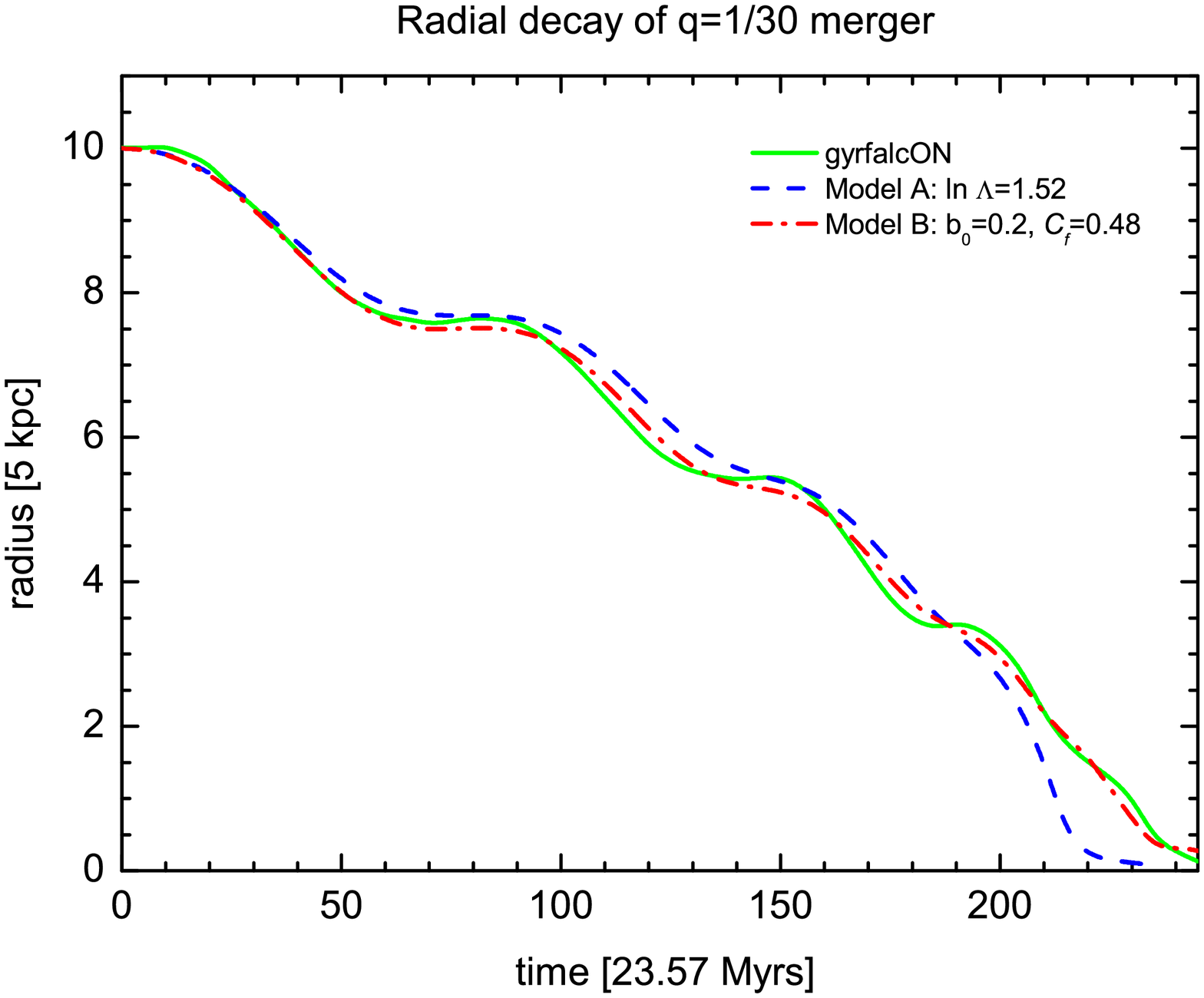} % {m030_A_and_B.eps}
\caption{Radial decay for an isothermal satellite within an isothermal halo for a mass ratio of $q=1/30$: Self-consistent reference model generated with \emph{gyrfalcON} (\textbf{green solid line}); best model A (\textbf{blue dashed line}) and best model B (\textbf{red dash-dotted line}).}
\label{petsch_fig_1}
\end{figure}

\begin{figure} \centering
\includegraphics[width=6cm] {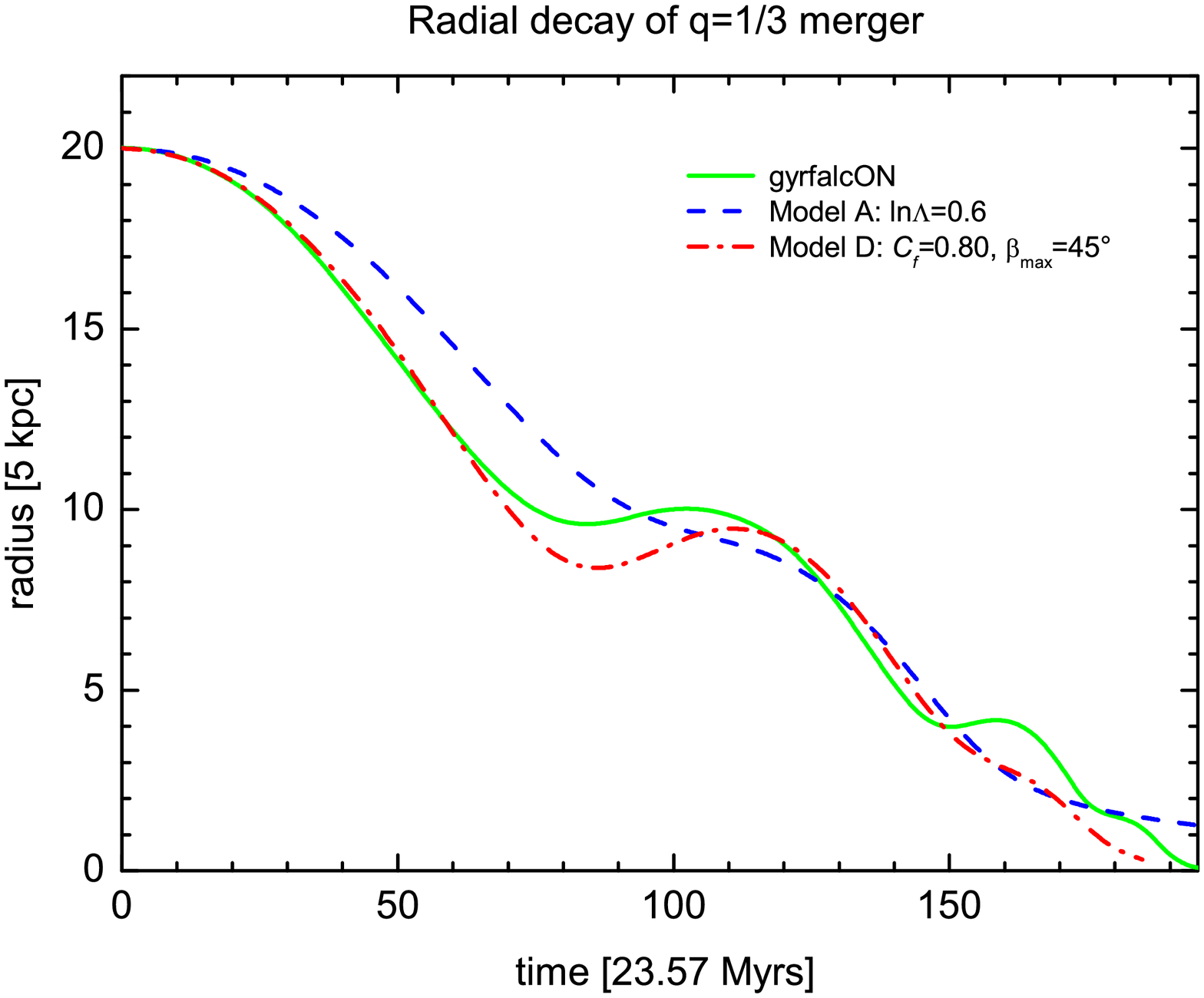} % {m003_A_and_D.eps}
\caption{Same as Fig.\,\ref{petsch_fig_1}, but for a mass ratio of $q=1/3$: Self-consistent model (\textbf{green solid line}); best model A (\textbf{blue dashed line}) and best model D (\textbf{red dash-dotted line}).}
\label{petsch_fig_2}
\end{figure}

\subsection{Isothermal satellite}
We mainly tested the merging of isothermal satellites into isothermal halos. In total more than 250\,000 restricted \textit{N-}body models were compared to 20 self-consistent reference models (with different mass ratios). Here we present two examples for initially circular orbits, but different mass ratios, i.e.\,$q=1/30 $ and $q=1/3$. The first example describes a low mass satellite. In that case all model approaches described in Sec.\,\ref{petsch_sec_method} are able to reproduce the radial decay of the satellite. In Fig.\,\ref{petsch_fig_1} we present the best fits for models with constant Coulomb logarithm (model A) and for a distance-dependent one (model B). The latter is superior because it is able to reproduce the complete merging process with a deviation\footnote{$\delta_d(t)$ is derived by integrating the quadratic difference between the compared radial decay curves over time $t$.} of $\delta_d(230)=1.8 \cdot 10^{-2}$. This example already shows the limitations of a constant Coulomb logarithm, the innermost part of the merging sequence occurs to quickly.
%Our results show, that it is possible to use dynamical friction to improve the restricted \textit{N-}body code and remodel the radial decay over a large range of mass ratios. 

The second example was done with a larger satellite mass (one third of the halo mass). In that case a constant Coulomb logarithm is not able to reproduce the merging process (Fig.\,\ref{petsch_fig_2}). It either leads to a large underestimation of the merging time or to a different behaviour of the radial decay (as shown there), resulting in a deviation of $\delta_d (185)=6.8 \cdot 10^{-2}$. Nevertheless, we were able to improve the models by using a mass-dependent Coulomb logarithm (model D). With this approach we could remodel the radial decay for one revolution with a deviation of $\delta_d (t_\mathrm{revolution}) \lesssim 1.5 \cdot 10^{-2}$ and $\delta_d (t_\mathrm{merge}) = 2.8 \cdot 10^{-2}$ for the complete merging.
%The merging process is strongly affected by feedback from the disturbed structures, that can be found in the self-consistent models. 

\subsection{Disk satellite}
\begin{figure} \centering
  \subfigure{\includegraphics[height=3.2cm]{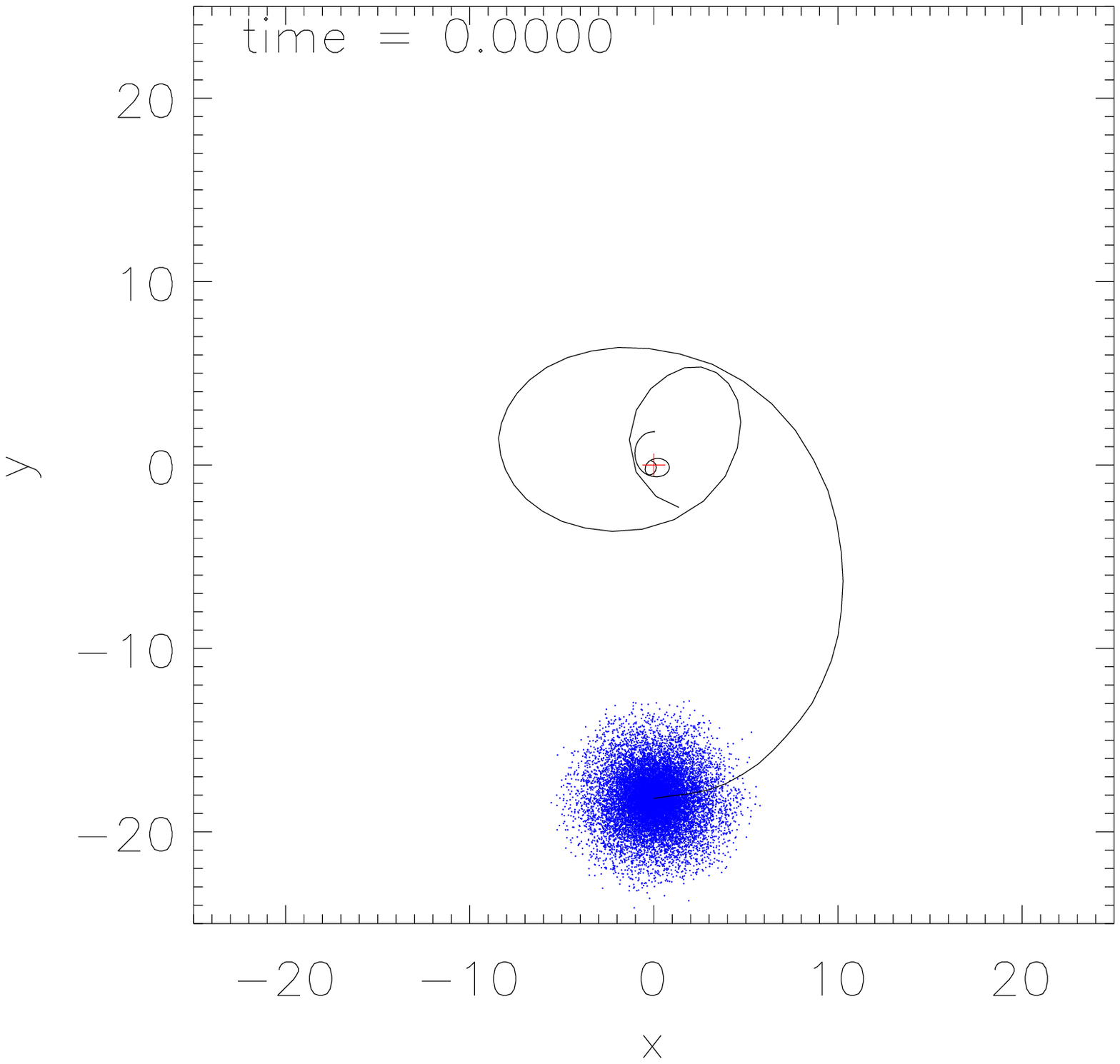}} %{d3aout001.eps}}
  \hspace{0cm}
  \subfigure{\includegraphics[height=3.2cm]{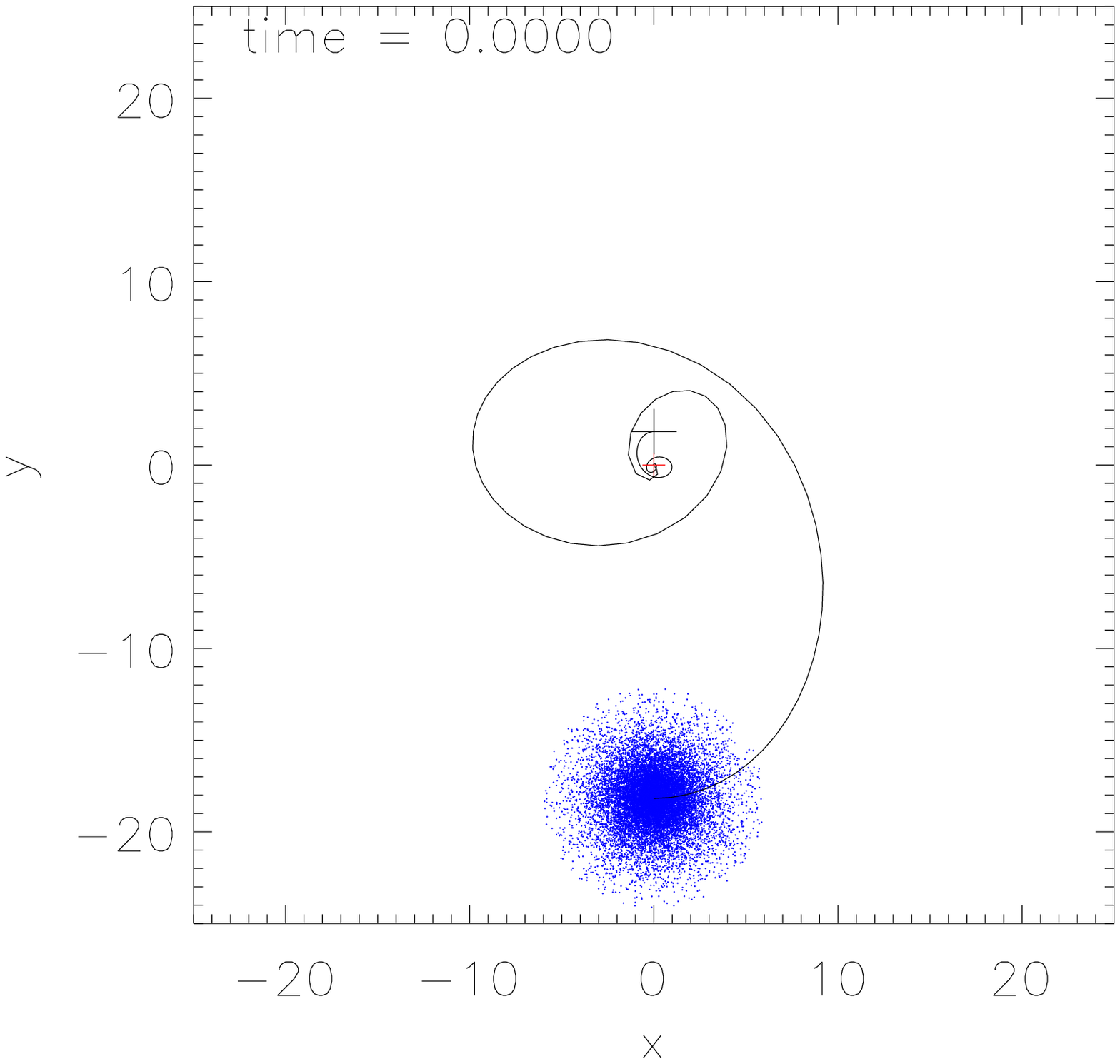}}\\%{10.out001.eps}}\\
  \vspace{0cm}
  \subfigure{\includegraphics[height=3.2cm]{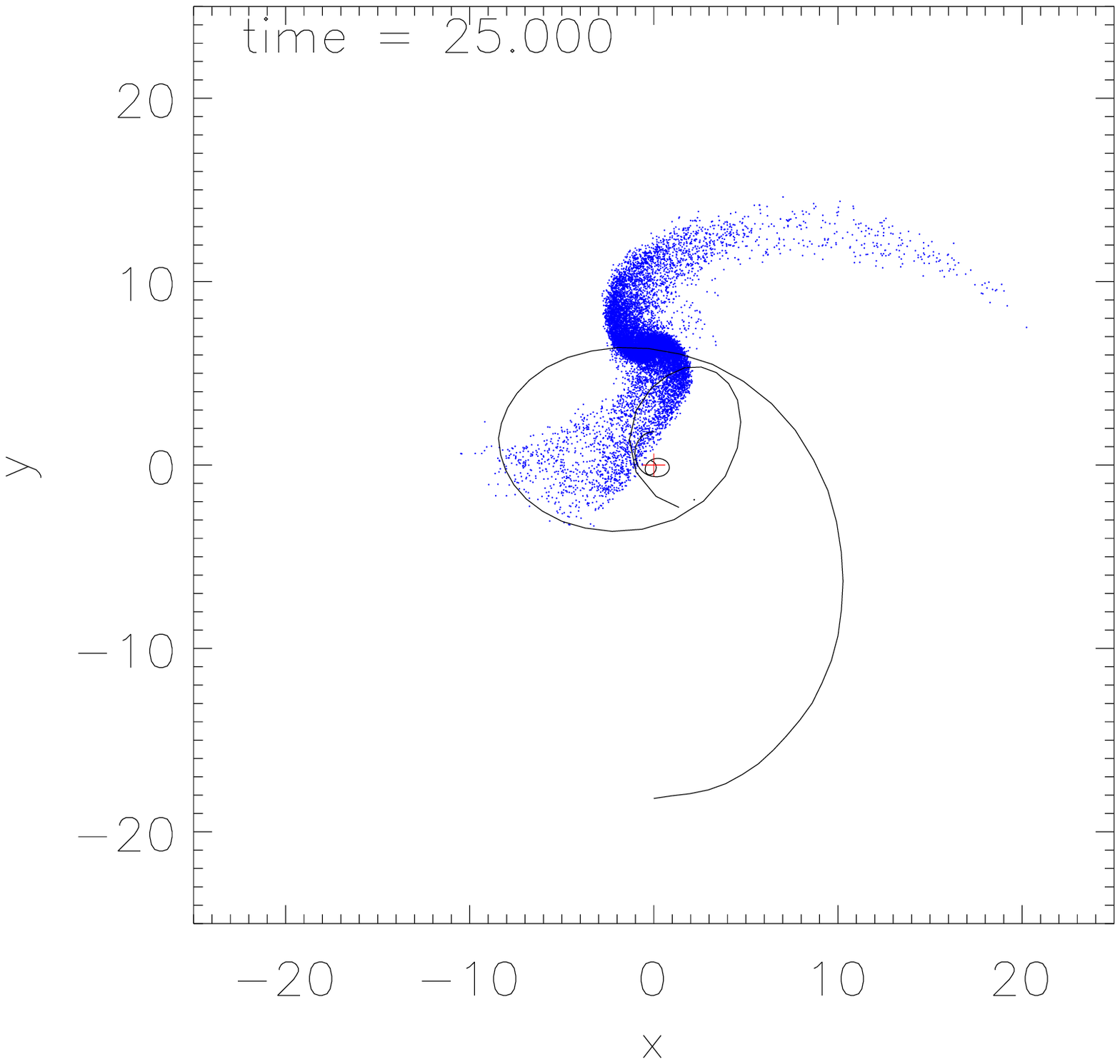}} %{d3aout006.eps}}
  \hspace{0cm}
  \subfigure{\includegraphics[height=3.2cm]{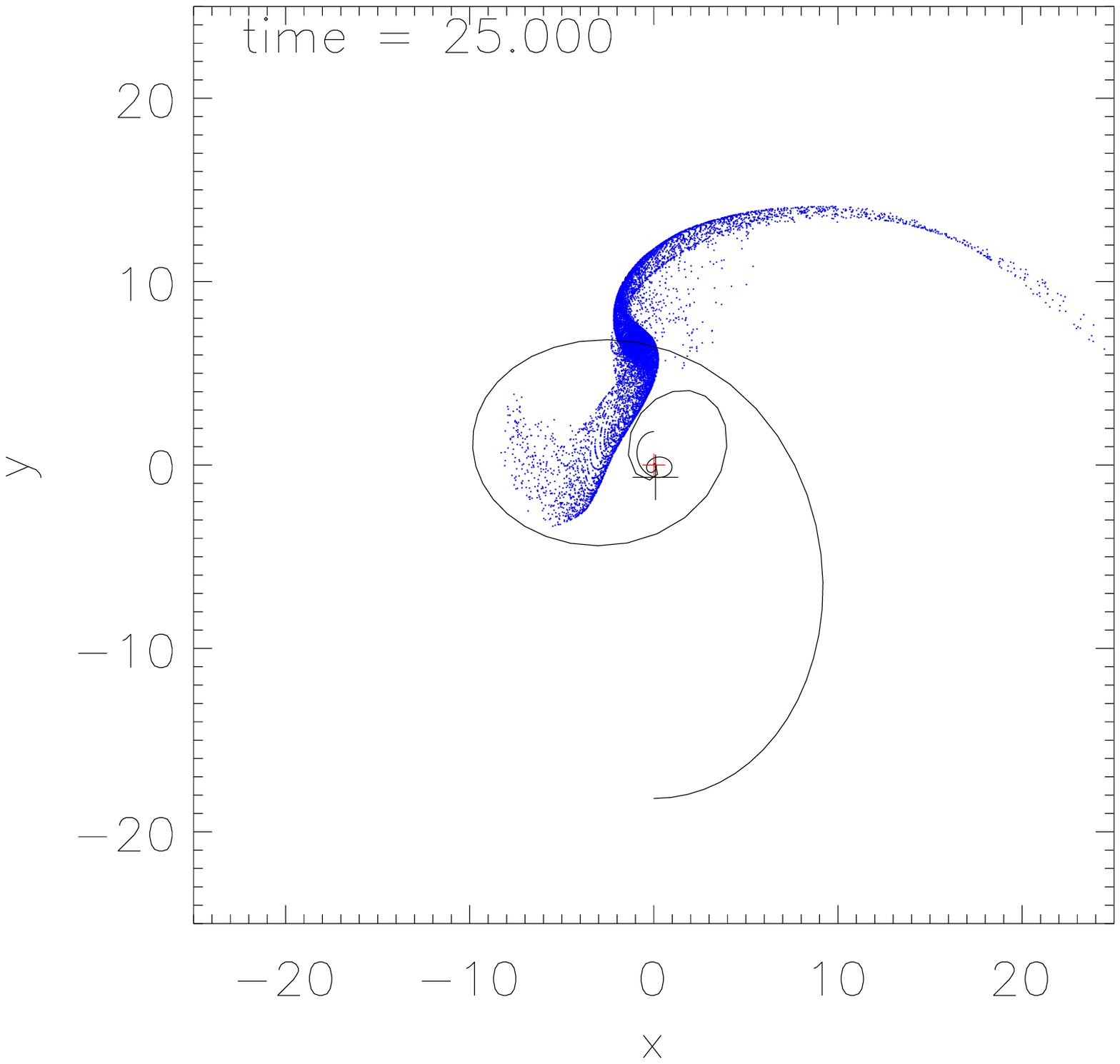}} %{10.out006.eps}} \\
  \vspace{0cm}
  \subfigure{\includegraphics[height=3.2cm]{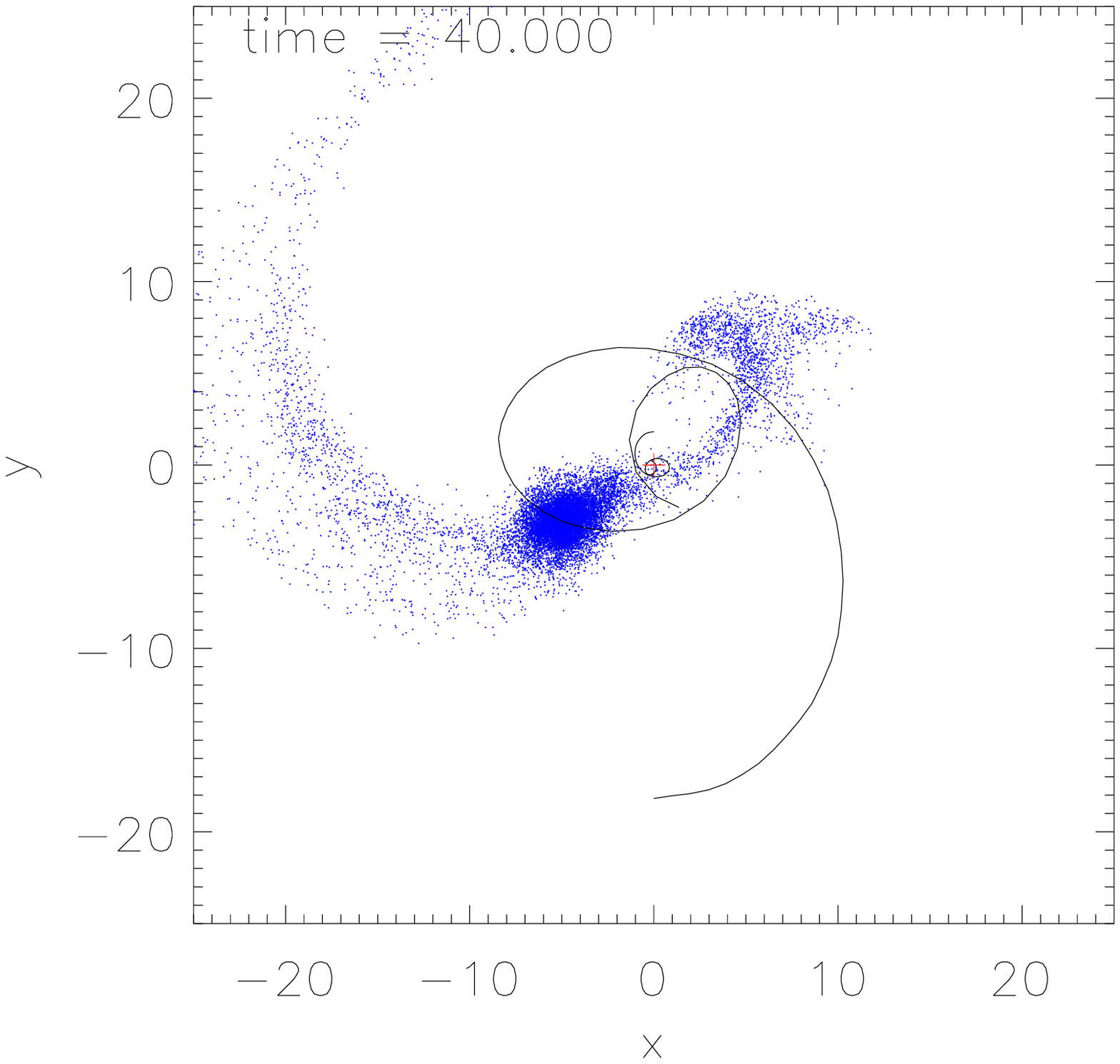}} %{d3aout009.eps}}
  \hspace{0cm}
  \subfigure{\includegraphics[height=3.2cm]{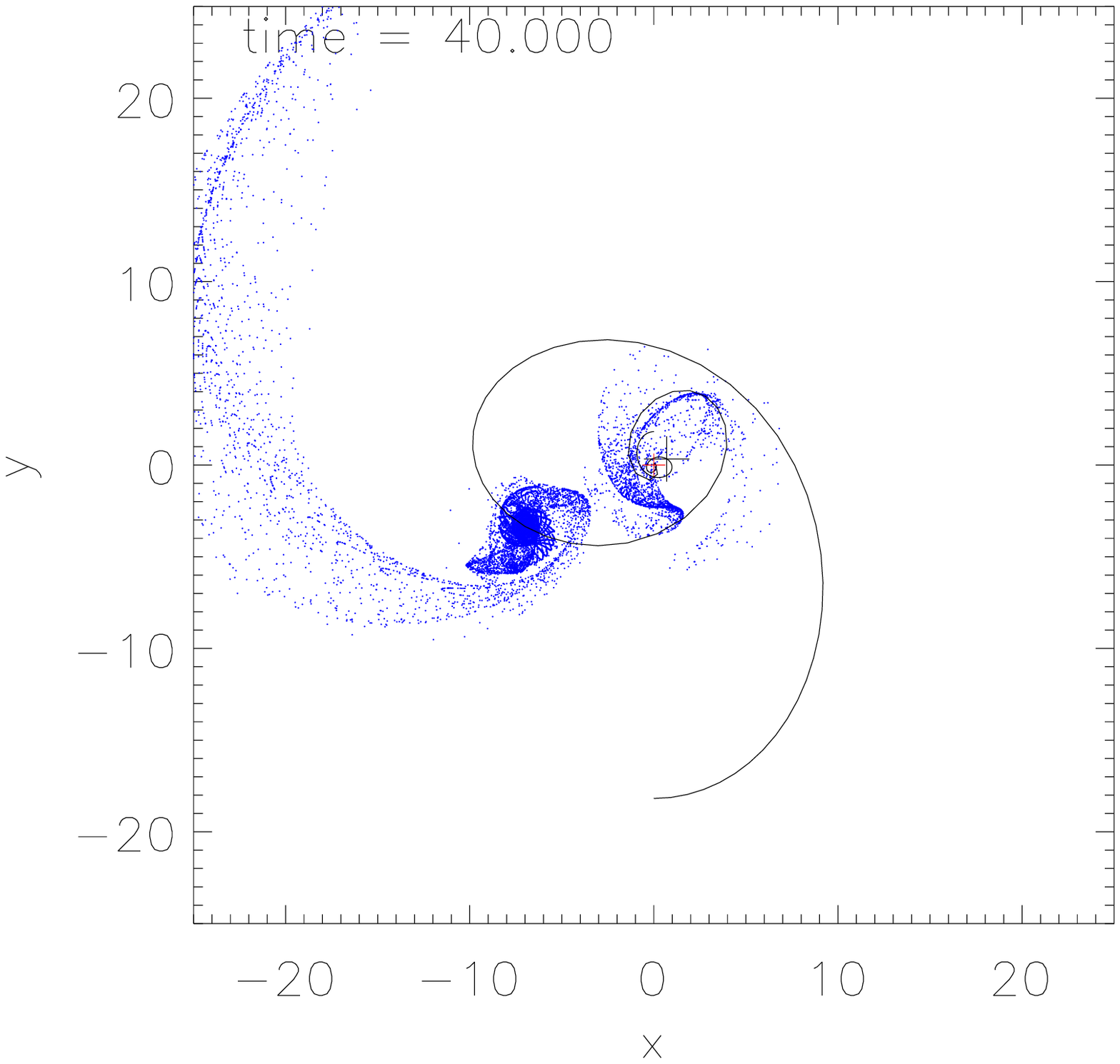}} %{10.out009.eps}}
  \caption{Model of a disk galaxy merging into an isothermal halo. Comparison between a self-consistent reference model (\textbf{left}) and an improved restricted \textit{N-}body model (\textbf{right}). For the self-consistent model only disk particles are shown (16\,000). The restricted model used 16\,129 test particles. The initial distribution was set to meet optically the initial reference model. Merging was completed at $\mathrm{time}=60.0$\,TU or $1.7$\,Gyrs. CPU time was $36$\,min for the self-consistent and $5.6$\,sec for the restricted model.}
\label{petsch_fig_3}
\end{figure}

We also carried out simulations using a disk-like satellite merging with an isothermal halo. A self-consistent model was set up with \textit{mkkd95} (Kuijken \& Dubinski 1995) and integrated with \textit{gyrfalcON}. The details of the model parameters can be found in Table \ref{petsch_tab_1}. The determination of the orbital decay was done in the same manner than for the isothermal satellites. As we could already use our results from the isothermal satellites, we only needed to carry out a few tenth of simulations. We also compared the location of the disk particles, i.e.\,the observables. Fig.\,\ref{petsch_fig_3} shows the good match of the radial decay and the comparable formation of the trailing tidal arm.  Minor mismatches like the distribution of particles in the trailing arm might be explained by the different initial setup of the disk. However, the leading arm of the disk galaxy could not be reproduced well. This shows, that we need to be careful about predictions for the innermost regions of the merger, derived from our models.

\begin{table}[htbp] 
	\centering
		\caption{Properties of the disk-merger test.}
\vspace{0.3cm}
	\begin{tabular} {lll}
		\hline \hline
	  \textbf{description} & \textbf{self-consistent} & \textbf{improved restricted} \\
		\hline
		Host-halo:&isoth., 65\,536 part. & isoth., static, analytic\\
 	        Disk galaxy:&self-consistent &restricted\\
 	  \ \ \ - halo:&12\,000 particles&static, analytic \\
 	           & $r_{\mathrm{halo}}\approx 10$\,kpc&$r_{\mathrm{halo}}=6$\,kpc \\
		\ \ \ - disk:	&16\,000 part. & 16\,129 part. (100 rings) \\
		integration:&\textit{gyrfalcON}&\textit{restricted N-body}\\
		CPU-time: & $36$\,min & $5.6$\,sec (incl. setup) \\
		Dyn. friction:&self-consistent&according equation (\ref{petsch_eq_6})\\
		&&$\beta=\pi/4$, $C_f=0.89$ \\
		\hline			
	\end{tabular}

  \label{petsch_tab_1}
\end{table}

\subsection{Genetic Algorithm run}
\begin{figure} \centering
\includegraphics[width=6cm]{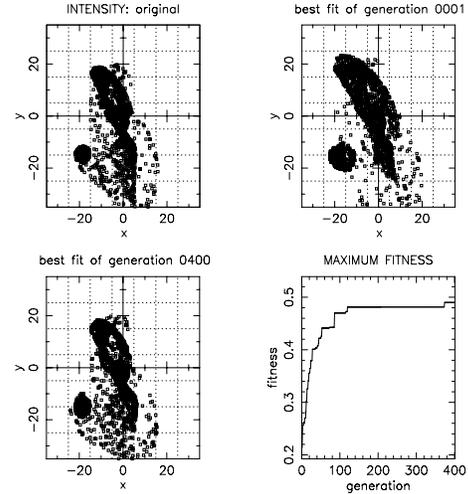} %{ga_tot_small_bb130.eps}
\caption{Result of a Genetic Algorithm run with MINGA: Particle distribution of the reference model shown in the \textbf{upper left} panel. The \textbf{upper right} panel shows the best model after generation 1. \textbf{Lower left}: best model found by the GA after 400 generations with 100 individuals each. \textbf{Lower right}: evolution of the fitness (defines how the original particle distribution is met) over 400 generations.}
\label{petsch_fig_4}
\end{figure}

The last result we want to present is a complete GA run with the implementation of dynamical friction to improve the restricted \textit{N-}body code of \textsc{Minga}. The GA was provided with a reference model (representing a real observation), a merger of two disc galaxies with a mass ratio $q=1/3$. Eight free parameters were selected -- see Table \ref{petsch_tab_2}. We have used our mass-dependent Coulomb logarithm of Eq.\,(\ref{petsch_eq_6}), where the strength of the dynamical friction $C_f$ was one of the free parameters. The results are shown in Fig.\,\ref{petsch_fig_4}. Most of the parameters were recovered with errors of less than $10\%$ cf.\,also Table \ref{petsch_tab_2}.

\begin{table}[htbp]
	\centering
		\caption{Interaction parameters that should be recovered by the GA. Parameter name, value for the reference model and provided limits are listed in columns 1 to 4. The recovered parameters can be found in column 5 and their relative errors in column 6.}

\vspace{0.3cm}
		\begin{tabular}{l|rrrrr} 
		\hline \hline
		\textbf{name} & \textbf{input} 	& \multicolumn{2}{c}{\textbf{limits}} & \textbf{recovered} &\textbf{rel. error}  \\
		\hline
    \textbf{$M_{\mathrm{halo},2}$} & 1.80& 0.54 & 2.70 & 1.70 & 5.7 $10^{-2}$ \\
    \textbf{$r_{\mathrm{disk},1}$} & 20.00 & 10.00 & 30.00 & 19.04& 4.8 $10^{-2}$ \\
    \textbf{$r_{\mathrm{disk},2}$} & 3.00 & 1.00 &10.00 & 3.23& 7.5 $10^{-2}$   \\
    \textbf{$r_{\mathrm{halo},2}$} & 3.00 & 1.00 & 10.00 & 3.02 & 7.7 $10^{-3}$\\
    \textbf{$\Delta z$} & 0.00 & -2.00 & 2.00 & -0.83& --- \\
    \textbf{$\Delta v_x$} & -0.1845 & -1.00 & 0.00 & -0.1500& 1.9 $10^{-1}$ \\
    \textbf{$\Delta v_y$} & 0.0537 & 0.00 & 1.00 & 0.0650& 2.1 $10^{-1}$ \\
    \textbf{$C_f$} & 0.50 & 0.10 & 1.00 & 0.46 & 8.6 $10^{-2}$ \\
		\hline
		\end{tabular}

	\label{petsch_tab_2}
\end{table}

%\begin{figure*}[t]
%\begin{center}
%\begin{picture}(400,5){
%\put(-120,140){\epsfig{file=d3aout001.eps, width=0.5\textwidth} } 
%\put(-120,-20){\epsfig{file=d3aout006.eps, width=0.5\textwidth} } 
%\put(-120,-180){\epsfig{file=d3aout009.eps, width=0.5\textwidth} } 
%\put(120,140){\epsfig{file=10.out001.eps, width=0.5\textwidth} } 
%\put(120,-20){\epsfig{file=10.out006.eps, width=0.5\textwidth} } 
%\put(120,-180){\epsfig{file=10.out009.eps, width=0.5\textwidth} } 
%}
%\end{picture} 
%\end{center}
%%\caption{ This is lovely, too}  
%\end{figure*} 

\section{Conclusions}
We have improved the restricted \textit{N-}body code by introducing dynamical friction. We varied the determination of the Coulomb logarithm as well as the strength and direction of the friction force. We compared our models to self-con\-sis\-tent simulations in order to find the best parameterisation. We have shown, that radial decays of mergers up to a mass ratio of $q=1/30$ can be reliable reproduced by using a constant or distance-dependent Coulomb logarithm. With the introduction of more sophisticated descriptions like a mass- and distance-dependent Coulomb logarithm, we were able to remodel radial decays for mergers up to a mass ratio of $q=1/3$. For these models it was also essential to use an orientation correction of the friction force. These improvements now account for a finite system with a density gradient. However, for equal mass mergers, we were not able to reproduce the orbital decay. Other neglected effects like mass loss might be the reason for failing remodeling. A few recent tests including mass loss already show promising results, so we might be able to improve the restricted \textit{N-}body code, again.

\acknowledgements
It is a pleasure to thank Peter Teuben for supplying the NEMO \emph{N-}body package and Walter Dehnen for providing the \emph{gyrfalcON} tree-code. This work was supported by the German Science Foundation (DFG) under the grant TH 511/9-1, which is part of the DFG priority program 1177.
% The authors are grateful to Chris Boily and his collaborators for organising the GSD2008 meeting in Strasbourg.

%\appendix

%\section{This is the title of the first appendix}


\begin{thebibliography}{}
  \bibitem{bin1} Binney, J., Tremaine, S.: 1987, \textit{Galactic Dynamics} (Princeton Univ. Press)
  \bibitem{chan1} Chandrasekhar, S.: 1942, \textit{Principles of stellar dynamics} (Dover: New York, 1960)
  \bibitem{char1} Charbonneau, P.: 1995, ApJS 101, 309
  \bibitem{deh1} Dehnen, W.: 2000, ApJ 536, L39
  \bibitem{gerd1} Gerds, Ch.: 2001, Diplomarbeit, Univ. Kiel
  \bibitem{hash1} Hashimoto, Y., Funato, Y., Makino, J.: 2003, ApJ 582, 196
  \bibitem{ji1} Jiang, C.Y., Jing, Y.P., Faltenbacher, A. et al.: 2008, ApJ 675, 1095
  \bibitem{ju1} Just, A.,  Pe\~{n}arrubia, J.: 2005, A\&A 431, 861
  \bibitem{kui1} Kuijken, K., Dubinski, J.: 1995, MNRAS 277, 1341
  \bibitem{mak1} Makino, J., Fukushige, T. et al.: 2003, PASJ 55, 1163
  \bibitem{pfle1} Pfleiderer, J., Siedentopf, H.: 1961, Zeitschrift f. Astrophysik 51, 201
  \bibitem{spinn1} Spinnato, P.F., Fellhauer, M., Portegies Zwart, S.F.: 2003, MNRAS 344, 22
  \bibitem{sugi1} Sugimoto, D., Chikada, Y., Makino, J. et al.: 1990, Nature 345, 33
  \bibitem{thei1} Theis, Ch.: 1999, Reviews in Modern Astronomy, Vol.\,12, 309
  \bibitem{thei2} Theis, Ch., Kohle, S.: 2001, A\&A 370, 365
  \bibitem{thei3} Theis, Ch.: 2004, IAU Symp. 220, 461
  \bibitem{too1} Toomre, A., Toomre, J.: 1972, ApJ 178, 623 (TT72)
  \bibitem{wah1} Wahde, M.: 1998, A\&AS 132, 417
\end{thebibliography}
\end{document}